\documentclass{aastex}          
\usepackage{spr-astr-addons}    
\usepackage{epsfig}
\usepackage{graphicx}
\usepackage{color}

\begin{document}
%
\title{Observation of standing kink waves in solar spicules}

\shorttitle{Standing kink waves in solar spicules}
\shortauthors{Ebadi et al.}

\author{H.~Ebadi\altaffilmark{1}}
\affil{Astrophysics Department, Physics Faculty,
University of Tabriz, Tabriz, Iran\\
e-mail: \textcolor{blue}{hosseinebadi@tabrizu.ac.ir}}
\and
\author{T.~V.~Zaqarashvili\altaffilmark{2}}
\affil{Space Research Institute, Austrian Academy of Sciences,\\ Schmiedlstrasse 6, A-8042 Graz, Austria
}
\and
\author{I.~Zhelyazkov}
\affil{Faculty of Physics, Sofia University, 5 James Bourchier Blvd., BG-1164 Sofia, Bulgaria}

\altaffiltext{1}{Research Institute for Astronomy and Astrophysics of Maragha,
Maragha 55134-441, Iran.}
\altaffiltext{2}{Abastumani Astrophysical Observatory at Ilia State University, 2 University Street, GE-0162 Tbilisi, Georgia}

\begin{abstract}
We analyze the time series of \mbox{Ca\,\textsc{ii}} H-line obtained from \emph{Hinode}/SOT on the solar limb.  The time-distance analysis shows that the axis of spicule undergos quasi-periodic transverse displacement at different heights from the photosphere.  The mean period of transverse displacement is $\sim\!\!180$ s and the mean amplitude is $1$ arc\,sec.  Then, we solve the dispersion relation of magnetic tube waves and plot the dispersion curves with upward steady flows.  The theoretical analysis shows that the observed oscillation may correspond to the fundamental harmonic of standing kink waves.
\end{abstract}


\section{Introduction}
\label{sec:intro}

Observation of oscillations in solar spicules may be used as an indirect evidence of energy transport from the photosphere towards the corona.  Transverse motion of spicule axis can be observed by both, spectroscopic and imaging observations.  The periodic Doppler shift of spectral lines have been observed from ground based coronagraphs \citep{nik67,Kukh2006,Tem2007}.  But Doppler shift oscillations with period of $\sim\!\!5$ min also have been observed on the SOlar and Heliospheric Observatory (\emph{SOHO}) by \citet{xia2005}.  Direct periodic displacement of spicule axes have been found by imaging observations on Optical Solar Telescope (SOT) on \emph{Hinode\/} \citep{De2007,Kim2008,he2009}.  The torsional Alfv\'en waves were reported recently in the context of a flux tube connecting the photosphere and the chromosphere as periodic variation of spectral line width \citep{jes09}.

The observed transverse oscillations of spicule axes were interpreted by kink \citep{nik67,Kukh2006,Tem2007,Kim2008} and Alfv\'{e}n \citep{De2007} waves.  All spicule oscillations events are summarized in a recent review by \citet{Tem2009}.
They suggested that the observed oscillation periods can be formally divided in two groups: those with shorter periods ($<\!\!2$ min) and those with longer periods ($\geqslant \!\!2$ min) \citep{Tem2009}.  The most frequently observed oscillations lie in the period ranges of $3$--$7$ min and $50$--$110$ s.

Spicule seismology, which means the determination of spicule properties from observed oscillations and was originally suggested by \citet{Tem2007}, has been significantly developed during last years \citep{Ajab2009,Verth2011,Ehsan2011}.

Spicules are almost $100$ times denser than surrounding coronal plasma \citep{bec68}, therefore they can be considered as cool magnetic tubes embedded in hot coronal plasma.  Wave propagation in a static magnetic cylinder was studied by Edwin and Roberts \citep{Edwin1983}.  They derived general dispersion relation of all possible wave modes in magnetic tubes.  Then the linear and non-linear MHD waves propagation in a cylindrical magnetic flux tube with axial steady flows have been also studied \citep{Terra2003}.  They show that steady flows change the treatments of propagating waves because of induced Doppler shifts.

In the present work, we study the observed oscillations in the solar spicules through the data obtained from \emph{Hinode}.  Then we model the oscillations as magnetohydrodynamic (MHD) waves in magnetic flux tubes.

\section{Observations and image processing}
\label{sec:observations}

We used a time series of \mbox{Ca\,\textsc{ii}} H-line ($396.86$ nm) obtained on 25 October
2008 during 03:20 to 03:25 UT by the Solar Optical Telescope onboard \emph{Hinode\/} \citep{Tsu2008}.  Note, that the \mbox{Ca\,\textsc{ii}} H-line observations of the same day have been used recently to study multi-component spicules by \citet{Ehsan2011}, but they used another time interval of these data.  The spatial resolution reaches $0.2$ arc\,sec ($150$ km) and the pixel size is $0.1082$ arc\,sec ($\sim\!\!80$ km) in the \mbox{Ca\,\textsc{ii}} H-line.  The time series has a cadence of $10$ seconds with an exposure time of $0.5$ seconds.  The position of $X$-center and $Y$-center of slot are, respectively, $0$ arc\,sec and $948$ arc\,sec, while, $X$-FOV and $Y$-FOV are $112$ arc\,sec and $56$ arc\,sec respectively.

We used the ``fgprep,'' ``fgrigidalign'' and ``madmax'' algorithms \citep{Koutchmy89} to reduce the images spikes and jitter effect and to align time series and to enhance the finest structures, respectively.

\begin{figure}[!h]
\epsscale{1.00}
\plotone{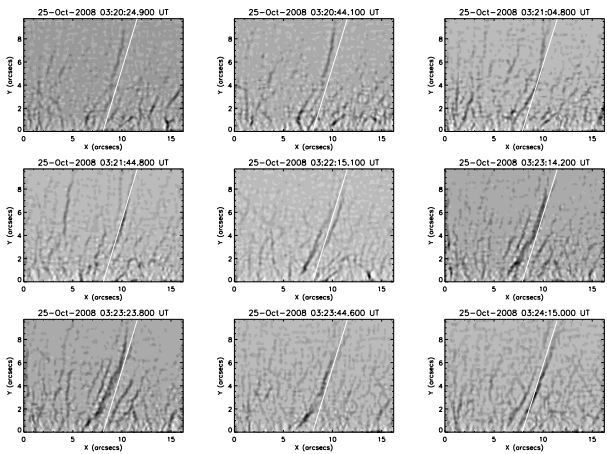}
\caption{9 images of time sequence in \mbox{Ca\,\textsc{ii}} H-line (left to right and top to bottom).  White diamonds indicate the spicule locations in time series.\label{fig1}}
\end{figure}

Figure~\ref{fig1} shows $9$ selected images of the time series, which consists of $27$ consecutive images.
We study the transverse motion of selected spicule (black narrow line) in respect with the hypothetic line (white line on the images), which is drown on the same place during whole time series.  The clear quasi-periodic transverse motion of the spicule axis is seen on the figure.

\begin{figure}[!h]
\epsscale{0.8}
\plotone{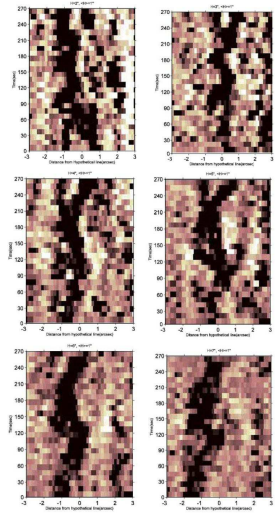}
\caption{Time slice images of time series shown in Fig.~\ref{fig1}.  The $x$-axis corresponds to the heliocentric coordinates measured from the white diamonds of Fig.~\ref{fig1}.\label{fig2}}
\end{figure}

We use the time slice diagrams in order to study the quasi-periodic motion of the spicule axis in detailed.  Figure~\ref{fig2} shows the time slice diagrams performed at $6$ different heights from the limb. Each cut is
obtained by averaging over $9$ pixels along the spicule axis, which corresponds to $1$ arc\,sec around each height.

\begin{figure}[!h]
\epsscale{1.00}
\plotone{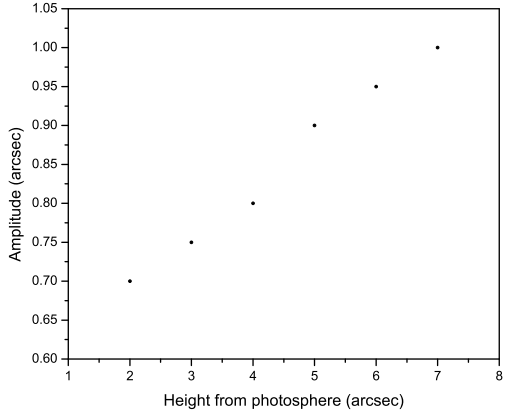}
\caption{The height variations of displacement amplitude of the spicule axis oscillation.\label{fig3}}
\end{figure}

The dark regions represent the spicule at particular height.  We clearly see that the spicule axis undergoes the transverse oscillation at each height with nearly same periodicity.  The oscillation period at each heights is estimated as $\sim\!\!\!180$ s. The oscillation amplitude is nearly $1$ arc\,sec, but slightly changes with height.  Figure~\ref{fig3} shows the height variation of the oscillation amplitude.  The
amplitude increases almost linearly with height and its slope is $0.06$.  The oscillations of spicules axis are closed to the standing pattern as we do not see any upward or downward propagation.

The density is almost homogeneous along the spicule axis \citep{bec68}, therefore the density scale height should be much longer that the spicule length.  Then the variation of oscillation amplitude with height can not be due to the decreased density.  Therefore, we assume that the height dependence of amplitude is due to the standing oscillations.  Then we may have that the oscillation amplitude is proportional to $\sin(kz)$, where $k$ is the oscillation wave number. This expression can be approximated in the long wavelength limit as $\simeq\!\!kz$.  Then, the oscillation wavelength is $\lambda \sim 1/0.06 \, \mathrm {arc \,sec}\sim 11500$ km.  This allows the estimation of the kink speed as $v_{k}=\lambda/T \sim 11500/180\,\mathrm {km\,s^{-1}}\sim 65$ km\,s$^{-1}$.  The estimated period and kink wave speed are in good agreement with the period and speed of fundamental harmonic of kink waves.

The oscillation amplitude and phase speed allows to estimate the energy flux, $F$, storied in the oscillations
\begin{equation}
\label{eq:flux}
F = \frac{1}{2}\rho v^{2} v_{k},
\end{equation}
where $\rho$, $v$ and $v_{k}$ are density, wave amplitude and phase speed respectively.  The density in spicules is $3\times10^{-10}$ kg\,m$^{-3}$ \citep{bec68}.  The wave velocity can be determined as $v=\xi_0/T_0$, where $\xi_0$ is the axis displacement estimated as $1$ arc\,sec and $T_0$ is the oscillation period estimated as $\sim\!\!180$ s.  We take the estimated phase speed as $v_{k} = 65$ km\,s$^{-1}$.  With these parameters the energy flux is estimated as $F$ = $1500$ J\,m$^{-2}$\,s$^{-1}$.  Total coronal energy losses in the quiet Sun is $3000$ J\,m$^{-2}$\,s$^{-1}$, therefore the energy flux storied in the oscillation is of the same order as the energy losses, but probably is not enough to fully compensate them.

\section{Theoretical modeling}
\label{sec:theory}

We use ideal MHD equations to model the propagation of waves in spicules.  The density seems almost uniform along almost whole length of spicules \citep{bec68}, therefore we ignore the effects of stratification due to gravity. The density inside spicules is almost $100$ times larger than outside, therefore we model them as magnetic flux tubes embedded in coronal magnetized plasma.  We consider a uniform vertical magnetic tube of radius $a$ with uniform magnetic field $B_{0}\hat{z}$ ($B_{\rm e}\hat{z}$) inside (outside) the tube.  The kinetic gas pressure and density inside (outside) of the tube are $p_{0}$ ($p_{\rm e}$) and $\rho_{0}$ ($\rho_{\rm e}$), respectively.  We also consider uniform steady flows inside (outside) the tube with velocity $U_{0}\hat{z}$ ($U_{\rm e}\hat{z}$).  The pressure balance condition at the tube boundary implies that
\begin{equation}
\label{eq:pres-balance}
p_{0}+\frac{B_{0}^2}{2\mu} = p_{\rm e} +
\frac{B_{\rm e}^2}{2\mu}.
\end{equation}
The densities inside and outside the tube are related as \citep{Edwin1983}
\begin{equation}
\label{eq:rhos-relat}
\frac{\rho_{0}}{\rho_{\rm e}} = \frac{2c_{\rm e}^{2} + \gamma v_{\rm Ae}^{2}}{2c_{0}^{2} + \gamma v_{\rm A0}^{2}},
\end{equation}
where $c_{0}=\sqrt{\gamma p_{0}/\rho_{0}}$ ($c_{\rm e}=\sqrt{\gamma p_{\rm e}/\rho_{\rm e}}$) and $v_{\rm A0} =
B_{0}/\sqrt{\mu\rho_{0}}$ ($v_{\rm Ae} = B_{\rm e}/\sqrt{\mu\rho_{\rm e}}$) are the sound and Alfv\'en speeds inside (outside) the tube, respectively. Here $\gamma$ is the ratio of specific heats and $\mu$ is the magnetic permeability.

Fourier transform of linearized MHD equations with the assumption that all the perturbed quantities are
$\propto\!\!\exp[\mathrm{i}(-\omega t + n \theta + kz)]$ and the continuity of the perturbed interface \citep{Chandra61} and the total pressure across the cylinder boundary $r = a$ yield the dispersion relation
\citep{Terra2003}
\begin{eqnarray}
\label{eq:disp-rel}
\rho_{\rm e}\left(\Omega_{\rm e}^{2} -
k^{2}v_{\rm Ae}^{2}\right)m_{0}\frac{I^{\prime}_{n}(m_{0}a)}{I_{n}(m_{0}a)} \nonumber \\
\nonumber \\
{}=\rho_{0}\left(\Omega_{0}^{2} - k^{2}v_{A0}^{2}\right)m_{\rm e}\frac{K^{\prime}_{n}(m_{\rm e}a)}{K_{n}(m_{\rm e}a)},
\end{eqnarray}
where $I_n$ and $K_n$ are modified Bessel functions of order $n$.  Here $m_{\alpha}$ and $\Omega_{\alpha}$ ($\alpha=0,\textrm{e}$) are given by the expressions
\begin{equation}
\label{eq:atten-coeffs}
m_{\alpha}^{2} = \frac{(k^{2}c_{\alpha}^{2} - \Omega_{\alpha}^{2})(k^{2}v_{\rm
A \alpha}^{2} - \Omega_{\alpha}^{2})}{(c_{\alpha}^{2} + v_{\rm A
\alpha}^{2})(k^{2}c_{\rm T \alpha}^{2} - \Omega_{\alpha}^{2})}
\end{equation}
and
\begin{equation}
\label{eq:omega}
\Omega_{\alpha} = \omega - k U_{\alpha},
\end{equation}
where
\begin{equation}
\label{eq:ct}
c_{\rm T \alpha} = \frac{c_{\alpha}v_{\rm A \alpha}}{\sqrt{c_{\alpha}^{2} + v_{\rm A \alpha}^{2}}}
\end{equation}
is the tube speed. This dispersion relation describes both, surface ($m_{0}^{2} > 0$) and body waves ($m_{0}^{2} < 0$).

\begin{figure}[!h]
\epsscale{0.90} \plotone{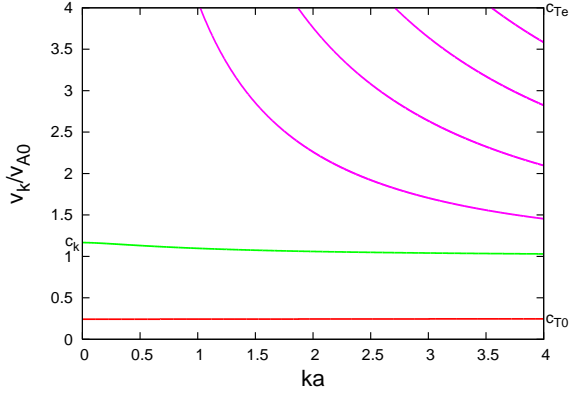} \caption{The phase speeds of the
kink modes (the curve with green color is fundamental kink mode)
under spicule conditions.
\label{fig4}}
\end{figure}

\begin{figure}[!h]
\epsscale{0.90} \plotone{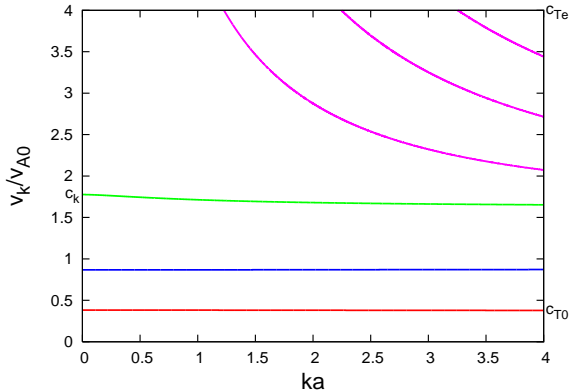} \caption{The same as in
Fig.~\ref{fig4} but for upward flow of $25$ km\,s$^{-1}$.\label{fig5}}
\end{figure}

We solved the dispersion relation for the values of $c_{\rm s0} = 10$ km\,s$^{-1}$, $c_{\rm se} = 262$ km\,s$^{-1}$, $v_{\rm A0} = 40$ km\,s$^{-1}$, $v_{\rm Ae} = 210$ km\,s$^{-1}$ and $\rho_{0}/\rho_{\rm e} = 0.02$.  With these data we have $c_{\rm k} = 42.5$ km\,s$^{-1}$, $c_{\rm T0} = 9.7$ km\,s$^{-1}$ and $c_{\rm Te} = 100.6$ km\,s$^{-1}$.  Figure~\ref{fig4} shows the dispersion diagram of kink waves under above
conditions in the absence of steady flows.  All speeds in the plots are normalized to $v_{\rm A0}$.  As $v_{\rm Ae} > v_{\rm A0}$ then the surface waves are absent \citep{Edwin1983}.  Figure~\ref{fig5} shows the
same dispersion diagram but in the case of steady flow $U_0 = 25$ km\,s$^{-1}$.  It is seen that phase speeds and consequently wave frequencies are shifted due to the steady flows.

Let us assume that the full length of spicule is $L$.  Then the wavelength of the fundamental standing mode is $\lambda = 4L$ and
thus the wave number of the fundamental mode is $k = 2\pi/\lambda = \pi/2L$.  The second and the third harmonics have the wave numbers $\pi/L$ and $2\pi/L$ , respectively.  The mean length of classical spicules varies from $5000$ to $9000$ km in H$_\alpha$ \citep{Tem2009}.  Then the periods of standing kink modes can be estimated from the dispersion diagram.  For the the spicule length of $3500$--$9000$ km we obtain the period of fundamental, second and third harmonics in the ranges of $170$--$320$ s, $85$--$160$ s and $40$--$80$ s, respectively.  The observed oscillation period of $\sim\!\!180$ s may correspond to the fundamental harmonic of standing kink waves.

\section{Discussion and conclusion}
\label{sec:concl}

We performed the analysis of Ca \begin{footnotesize}II\end{footnotesize} H-line time series at the solar limb obtained from \emph{Hinode}/SOT in order to uncover the oscillations in the solar spicules.  We concentrate on particular spicule and found that its axis undergos quasi-periodic transverse displacement about a hypothetic line.  The period of the transverse displacement is $\sim\!\!\!180$ s and the mean amplitude is $\sim\!\!1$ arc\,sec.  The same periodicity was found in Doppler shift oscillation by \citet{Tem2007} and \citet{De2007}, so the periodicity is probably common for spicules.

We model spicules as dense plasma jets along magnetic flux tube being injected from the chromosphere upwards into the hot corona.
Solving the wave dispersion relation in the presence of steady flows we conclude that the steady flows change the characteristics of wave propagation in a cylindrical magnetic flux tube. The calculated periods of fundamental, second and third harmonics of standing kink modes with an upward flow of $25$ km\,s$^{-1}$ are in the ranges of $170$--$320$ s, $85$--$160$ s and $40$--$80$ s, respectively, for the spicule length of $3500$--$9000$ km.

Therefore, the observed quasi-periodic displacement of spicule axis can be caused due to fundamental standing mode of kink waves.
The energy flux storied in the oscillation is estimated as $150$ J\,m$^{-2}$\,s$^{-1}$, which is of the order of coronal energy losses in quiet Sun regions.

\acknowledgments
The authors are grateful to the \mbox{\emph{Hinode}} Team for providing the observational data.  \mbox{\emph{Hinode\/}} is a Japanese mission developed and lunched by ISAS/JAXA, with NAOJ as domestic partner and NASA and STFC(UK) as international partners.  Image processing Mad-Max program was provided by Prof.~O.~Koutchmy.  This work has been partly supported by RIAAM.  The work of T.Z.\ was supported by the Austrian Fonds zur F\"orderung der wissenschaftlichen Forschung (projects P21197-N16) and from the Georgian National Science Foundation (under Grant GNSF/ST09/4-310).

\makeatletter
\let\clear@thebibliography@page=\relax
\makeatother

\end{document}